\documentclass{jpsj-suppl}
\usepackage{txfonts} 
\usepackage{bm}
\newcommand{\bra}[1]{\langle \, #1 \, |}
\newcommand{\ket}[1]{| \, #1 \, \rangle}
\newcommand{\kket}[1]{\, #1 \, \rangle}
\newcommand{\physdim}[1]{\hspace{1ex} \mathrm{#1}}

\title{Model-independent determination of the compositeness of near-threshold quasibound states}

\author{Yuki \textsc{Kamiya} and Tetsuo \textsc{Hyodo}}

\inst{Yukawa Institute for Theoretical Physics, Kyoto University Kitashirakawa Oiwakecho, Sakyo-ku, Kyoto 606-8502 Japan, \\
}

\email{yuki.kamiya@yukawa.kyoto-u.ac.jp}

\recdate{June 23, 2011}

\abst{
We study the compositeness of near-threshold states to clarify the internal structure of exotic hadron candidates. 
Within the framework of effective field theory, we extend the Weinberg's weak-binding relation to include the nearby CDD (Castillejo-Dalitz-Dyson) pole contribution with the help of the Pad\'e approximant. 
Finally, using the extended relation, we conclude that the CDD pole contribution to the $\Lambda(1405)$ baryon in the $\bar{K}N$ amplitude is negligible.
}
\kword{hadron structure, compositeness}

\begin{document}
\maketitle
\section{Introduction}

Motivated by the recent discoveries of many candidates of the exotic hadrons~\cite{Swanson:2006st,Brambilla:2010cs}, which are not assigned to the $qqq$ baryon state nor the $q\bar{q}$ meson state, the study of the internal structure is an essential subject in hadron physics. 
While the much experimental and theoretical efforts have been paid for this topic, there are many hadrons whose structure is not clarified.

The powerful method  to discuss the structure of shallow bound states directly from the experimental observables is established by Weinberg~\cite{Weinberg:1965zz}. %
He has shown that the deuteron is a proton-neutron composite system deriving the relations between the threshold parameters (the scattering length $a_0$ and the effective range $r_e$) and the normalization of the deuteron wave function. The internal structure of the deuteron is characterized by the compositeness $X$ which is defined as the probability to find the composite state in the physical bound state using the wave function of the bound state~\cite{Hyodo:2013nka,Sekihara:2014kya}. The weak-binding relation is given by %
\begin{equation}
	a_{0} = R\left\{ \frac{2X}{1+X} + {\mathcal O}\left(\frac{R_{\mathrm{typ}}}{R}\right)\right\},\quad r_e= -R\left\{\frac{1-X}{X}+ {\mathcal O}\left(\frac{R_{\mathrm{typ}}}{R}\right)\right\}, \label{eq:comp-rel-bound}
\end{equation}
where  $\mu$ is the reduced mass and $R =1/\sqrt{2\mu B}$ is the length scale related to the binding energy $B$. %
$R_{\mathrm{typ}}$ is the typical length scale of the interaction of this system. These weak-binding relations tell us that if the binding energy of the state is small and satisfies $R_{\mathrm{typ}}/R\ll 1$, the compositeness can be calculated from the observables without using models. To study the structure of exotic hadrons, which are all unstable states, the weak-binding relation is generalized for an unstable state in Ref.~\cite{Kamiya:2015aea,Kamiya:2016oao} %

On the other hand, as already mentioned in Ref.~\cite{Weinberg:1965zz}, we cannot apply the weak-binding relations~\eqref{eq:comp-rel-bound} to the system which has a CDD (Castillejo-Dalitz-Dyson) pole~\cite{Castillejo:1955ed} near the threshold. %
The CDD pole is defined as the pole of the inverse amplitude $f^{-1}$. It is considered that the existence of the CDD pole reflects the contribution comes from outside the model space~\cite{Chew:1961}.
Recently, the relation between the CDD pole position and the compositeness is discussed in detail in Refs.~\cite{Baru:2010ww,Guo:2016wpy}. In this paper, we present the extension of the weak-binding relation taking the nearby CDD pole contribution.
All the contents in this paper is based on Ref.~\cite{Kamiya:2016oao}.

\section{Derivation of the weak-binding relation based on effective field theory}\label{Sec:Derivation}
We first consider the single-channel $s$-wave scattering with a shallow bound state. Our interest is focused on the low energy physics near the threshold. We analyze this system with the non-relativistic effective field theory (EFT) with contact interactions~\cite{Kaplan:1996nv,Braaten:2007nq}
\begin{align}
H_{\mathrm{free}} =&\int d\bm{r}
\biggl[\tfrac{1}{2 M} \mathbf{\nabla} \psi^\dagger \cdot\mathbf{\nabla} \psi +\tfrac{1}{2 m} \mathbf{\nabla} \phi^\dagger \cdot\mathbf{\nabla} \phi 
 + \tfrac{1}{2 M_{0}} \mathbf{\nabla}  B_0^\dagger \cdot{\mathbf \nabla} B_0 +  \nu_0 B_0^\dagger B_0
\biggr] ,\\
H_{\mathrm{int}} =& \int d\bm{r}
\left[
g_{0} \left( B_0^\dagger \phi\psi + \psi^\dagger\phi^\dagger B_0 \right) + \lambda_0 \psi^\dagger\phi^\dagger \phi\psi
\right].
\end{align}
Here we take $\hbar =1$. We consider that this EFT is applicable below the cutoff momentum $\Lambda$. The cutoff scale is related to the typical length scale of the interaction $R_{\mathrm{typ}}$ as $ \Lambda\sim 1/R_{\mathrm{typ}}$.
We consider that the full Hamiltonian has a discrete eigenstate $|B\rangle$ with the binding energy $B$ in the same quantum numbers with the two-body $\psi\phi$ system as $H\ket{B}=-B\ket{B}$.
From the phase symmetry of the Hamiltonian, we can show that the completeness relation in this sector is spanned by the eigenstates of $H_{\mathrm{free}}$: the scattering states $\ket{\bm{p}}=\tilde{\psi}^{\dag}(\bm{p})\tilde{\phi}^{\dag}(-\bm{p})/\sqrt{\mathcal{V}}\ket{0}$ and the discrete state $\ket{B_{0}}=\tilde{B}_{0}^{\dag}(\bm{0})/\sqrt{\mathcal{V}}\ket{0}$ with the creation operators $\tilde{\psi}^{\dag}(\bm{p})$, the vacuum $\ket{0}$, and $\mathcal{V}=(2\pi)^{3}\delta^{3}(\bm{0})$.
Thus the physical bound state $|B\rangle$ can be written as a linear combination of these states.
Now we define the compositeness $X$ (the elementariness $Z$) as the probability to find the scattering (discrete) state in the bound state;
\begin{eqnarray}
X \equiv \int \tfrac{d\bm{p}}{(2\pi)^{3}} |\bra{\bm{p}}\kket{B}|^{2},\quad Z \equiv |\bra{B_{0}}\kket{B}|^{2}.
\label{eq:normalization}
\end{eqnarray}
By normalizing the bound state as $\langle B|B\rangle = 1$, $X$ and $Z$ satisfy the important relations; $Z+X =1$ and $Z,X \in [0,1]$, which ensure the probabilistic interpretation.

In this formalism, it is known that the compositeness $X$ can be written as~\cite{Sekihara:2014kya,Kamiya:2016oao} 
\begin{eqnarray}
X = -g^2 G^\prime(-B) \label{eq:X-gG},
\end{eqnarray}
where $g^2 \equiv \lim_{E\rightarrow -B}(E+B)t(E)$ is the coupling constant between the scattering state and the bound state, $G(E)$ is the loop function normalized by the cutoff $\Lambda$.
To derive the weak-binding relation, assuming that the binding energy $B$ is sufficiently small, we express
$g^2$ and $G^\prime(-B)$ with the experimental observables.
In Ref.~\cite{Sekihara:2014kya}, it is shown that the $G^\prime(-B)$ can be expanded as 
\begin{eqnarray}
G^\prime(-B) =-\tfrac{\mu^2 R}{2\pi} \left[1+\mathcal{O}\left(\tfrac{R_{\mathrm{typ}}}{R}\right)\right] \label{eq:G-prime-exp}.
\end{eqnarray}
In this expression, the leading term is independent of $R_{\mathrm{typ}}$ and written solely by $B$.
To approximate the coupling constant $g^2$, we introduce the effective range expansion (ERE) for the inverse amplitude:
\begin{align}
f^{-1}(p) =p\cot\delta -ip,\quad p\cot\delta =-\tfrac{1}{a_0} +\tfrac{r_e}{2} p^2 +\mathcal{O}(R_{\mathrm{eff}}^3p^4) \label{eq:eff-exp},
\end{align}
where $R_{\mathrm{eff}}$ is the length scale characterizing this expansion.
With Eq,~\eqref{eq:eff-exp}, $g^2$ can be expressed as
\begin{align}
g^{2} &= \tfrac{2\pi}{\mu^2} (R-r_e +R\mathcal{O}((R_{\mathrm{eff}}/R)^3))^{-1} \label{eq:g-exp}
\end{align}
By substituting Eqs.~\eqref{eq:G-prime-exp} and \eqref{eq:g-exp} into Eq.~\eqref{eq:X-gG}, the weak-binding relation is obtained:
\begin{align}
X    &= \left[1-\frac{r_e}{R} +\mathcal{O}\left(\left(\tfrac{R_{\mathrm{eff}}}{R}\right)^3\right)\right]^{-1}\left[1 + \mathcal{O}\left(\tfrac{R_{\mathrm{typ}}}{R}\right)\right] \label{eq:comp-rel-re-R} .
\end{align}
Here we notice that we introduce two assumptions:
\begin{itemize}
\item[(i)] $R$ is sufficiently smaller than the typical range scale of the interaction: $R_{\mathrm{typ}}/R \ll 1$,
\item[(ii)] the convergence region of ERE reaches the bound state pole: $(R_{\mathrm{eff}}/R)^3 \ll 1$.
\end{itemize}
The higher order term of $\mathcal{O}(R_{\mathrm{typ}}/R)$ comes from the assumption (i), which is related to the expansion of the derivative of the loop function~\eqref{eq:G-prime-exp}.
With the assumption (ii), the higher order term of $\mathcal{O}((R_{\mathrm{eff}}/R)^3)$ arises from the expansion of  $g^{-2}$. 
To simplify the expression~\eqref{eq:comp-rel-re-R} as Eq.~\eqref{eq:comp-rel-bound}, we need the stronger assumption  $R_{\mathrm{eff}} \lesssim R_{\mathrm{typ}}$ instead of the assumption (ii).
Without this assumption, two expansions are independent of each other and these higher order terms should be considered separately.

\section{Improvement of the weak-binding relation}\label{Sec:Improvement}
While the ERE is a general expression of the near-threshold amplitude, its convergence region does not always reach the eigenenergy.
The convergence region of the ERE is determined by the magnitude of $R_{\mathrm{eff}}$ in Eq.~\eqref{eq:eff-exp}.
When $R_{\mathrm{eff}}$ is large, the convergence region is limited to be small.
A large $R_{\mathrm{eff}}$ is sometimes caused by the nearby CDD (Castillejo-Dalitz-Dyson) pole~\cite{Castillejo:1955ed}.
Because the energy of the CDD pole $E_c$ is defined as the energy where the scattering amplitude vanishes,
the ERE converges only in the region $|E|<|E_c|$.
When $|E_c| \lesssim B$, the ERE is not valid at $E=-B$, and then we cannot use the weak-binding relation~\cite{Weinberg:1965zz}.
Because the ERE is introduced to describe $g^{-2}$ in Eq.~\eqref{eq:X-gG}, in the following two subsections, we extend the weak-binding relation by employing improved expressions of $g^{-2}$.

\subsection{Improvement by higher order terms in effective range expansion}\label{mod-ERE}
While we have approximated $p\cot\delta$ in Eq.~\eqref{eq:eff-exp} up to the $p^2$ term,
here we take the fourth order term of the expansion to improve the approximation: %
\begin{align}
p\cot\delta = -\tfrac{1}{a_0} + \tfrac{r_e}{2}p^2 + \tfrac{v}{4}p^4+\mathcal{O}(R_{\mathrm{eff}}^5p^6)\label{eq:mod-e.r.e}.
\end{align}
With Eqs.~\eqref{eq:mod-e.r.e} and \eqref{eq:g-exp}, the coupling constant $g^2$ can be expressed with $R$, $r_e$ and $v$. %
With these quantities, the compositeness is written as %
\begin{align}
X  =\left[1-\frac{r_e}{R}+\frac{v}{R^3} +\mathcal{O}\left(\left(\frac{R_{\mathrm{eff}}}{R}\right)^5\right)+\mathcal{O}\left(\frac{R_{\mathrm{typ}}}{R}\right)\right]^{-1}\label{eq:comp-rel-mod-ERE1}. 
\end{align}
When $v/R^3$ is order of $\mathcal{O}((R_{\mathrm{eff}}/R)^3)$, 
this expression reduces to Eq.~\eqref{eq:comp-rel-re-R}. %
We can improve the estimation of the compositeness by including the term of $v/R^3$.
By the condition of the bound state, this relation can be rewritten in terms of $R$, $a_0$ and $r_e$ as 
\begin{align}
X  =\left[\frac{4R}{a_0}+\frac{r_e}{R}-3 +\mathcal{O}\left(\left(\frac{R_{\mathrm{eff}}}{R}\right)^5\right)+\mathcal{O}\left(\frac{R_{\mathrm{typ}}}{R}\right)\right]^{-1}\label{eq:comp-rel-mod-ERE2}.
\end{align}
When $R$ satisfies $(R_{\mathrm{eff}}/R)^5\ll 1 $ and $R_{\mathrm{typ}}/R\ll 1$, we can neglect the higher order terms and calculate the compositeness from $a_0$, $r_e$ and $B$. 
In Eq.~\eqref{eq:comp-rel-mod-ERE2}, the contribution from the higher order terms of the effective range expansion is included in $R$ through the bound state condition. 

\subsection{Improvement by Pad\'e approximant}\label{subsec:Pade}

As mentioned above, the nearby CDD pole restrict the convergence region of the ERE. %
To include the CDD pole contribution to the weak-binding relation, here we use the Pad\'e approximant method to describe $p\cot\delta$: %
\begin{align}
p\cot\delta = \frac{b_0 + b_1 p^2}{1 + c_1p^2} + \mathcal{O}\left( R_{\mathrm{Pad\acute{e}}}^5p^6\right)\label{eq:Pade},
\end{align}
where $R_{\mathrm{Pad\acute{e}}}$ is the length scale characterizing this expansion.
The threshold parameters are related to this expansion as $a_0 = -b_0^{-1}$ and $r_e = 2(b_1-b_0c_1)$. %

Substituting Eq.~\eqref{eq:Pade} to Eq.~\eqref{eq:g-exp}, we obtain the expression of the coupling constant: %
\begin{align}
g^2 =& - \frac{2\pi p}{\mu^2}\left.\left\{\frac{2b_1p (1+c_1p^2 ) - 2c_1p(b_0 + b_1 p^2)   }{\left(1+c_1p^2 \right)^2} -i+\mathcal{O}\left(R_{\mathrm{Pad\acute{e}}}^5p^5 \right)\right\}^{-1}\right|_{p=i/R}.
\end{align}
We can replace the three independent quantities $b_0$, $b_1$ and $c_1$ by the threshold parameters $a_0$, $r_e$ and $R$ using the condition of the bound state.
Thus the extended weak-binding relation becomes
\begin{align}
X=\left[1 - \frac{4R(a_0-R)^2}{a_0^2r_e}+ \mathcal{O}\left(\left(\frac{R_{\mathrm{Pad\acute{e}}}}{R}\right)^5\right)+\mathcal{O}\left(\frac{R_{\mathrm{typ}}}{R}\right)\right]^{-1}\label{eq:comp-rel-Pade2}.
\end{align}
By neglecting the two higher order terms, we can calculate the compositeness with the three observables $a_0$, $r_e$ and $B$. %
With this extended relation, the nearby CDD pole contribution can be included in the determination of the compositeness owing to the employment of the Pad\'e approximant. %

\subsection{Application to $\Lambda(1405)$}\label{subsec:application}

$\Lambda(1405)$ is the negative parity excited baryon with spin $1/2$ and isospin $I=0$~\cite{Agashe:2014kda}.
$\Lambda(1405)$ couples to the $\bar{K}N$ channel in $s$ wave and eventually decays into the $\pi\Sigma$ channel. 
In the framework of the chiral SU(3) dynamics \cite{Ikeda:2011pi,Ikeda:2012au}, it is pointed out that the $\Lambda(1405)$ is well described by the two pole structure. %
We can study the $\bar{K}N$ compositeness of the higher pole state which lies near the $\bar{K}N$ threshold energy. %
Based on the weak-binding relation and some experimental analyses, without considering the CDD pole contribution, it is shown that $\Lambda(1405)$ is dominated by the $\bar{K}N$ composite component~\cite{Kamiya:2015aea,Kamiya:2016oao}. %

To investigate the CDD pole contribution to $\Lambda(1405)$, we calculate the compositeness with three weak-binding relations~\eqref{eq:comp-rel-bound}, \eqref{eq:comp-rel-mod-ERE2} and \eqref{eq:comp-rel-Pade2}, using the scattering length $a_0=1.39-i0.85\physdim{fm}$, the effective range $r_e=0.24-i0.05\physdim{fm}$ and the eigenenergy $E_h=-10-i26\physdim{MeV}$ determined from the amplitude in Ref.~\cite{Ikeda:2012au}.
Then the calculated values of $X_{\bar{K}N}$ are
$1.3+i0.1$ with Eq.~\eqref{eq:comp-rel-bound}, 
$1.3+i0.2$ with Eq.~\eqref{eq:comp-rel-mod-ERE2} 
and $1.4+i0.2$ with Eq.~\eqref{eq:comp-rel-Pade2}.
We can see that these values are close to each other. %
This means that the ERE converges well and the CDD pole contribution to this state can be neglected.

\section{Conclusion}
We have discussed the nearby CDD pole contribution to the weak-binding relation in the framework of the nonrelativistic effective field theory. %
In the new derivation of the weak-binding relation, we separate the expansion in terms of $R_{\mathrm{typ}}/R$ from that of $R_{\mathrm{eff}}/R$. %
This derivation clarifies the relation between the higher order terms in the two different expansions and evades the implicit assumption $R_{\mathrm{eff}} \lesssim R_{\mathrm{typ}}$ in the previous derivations.
Then by employing two different improvements of the approximation of the scattering amplitude, we show the extended weak-binding relation.
The first extended relation is obtained by taking the fourth order term in terms of $p$ in the effective range expansion.
The second improvement is accomplished by introducing the  Pad\'e approximant for the inverse scattering amplitude to include the nearby CDD pole contribution. %
Finally, we discuss the compositeness of the higher pole state of the $\Lambda(1405)$ using the extended weak-binding relation. %
 From the small difference of the compositeness calculated with the three weak-binding relations, we conclude that the effective range expansion converges well and the CDD pole contribution is negligible in the $\bar{K}N$ amplitude.%

\end{document}